\documentclass[aps,prb,preprint,groupedaddress]{revtex4-1}

\usepackage{color}
\usepackage[]{graphicx}
\usepackage[]{amsmath}  %>>> for AMS math formatting,
\usepackage[]{amssymb}

\begin{document}

%\preprint{}

\title{Magnetotunnelling in resonant tunnelling structures with spin-orbit interaction}


%\author{}
%\thanks{}
%\affiliation{}

\author{Goran Isi\'c}
\affiliation{School of Electronic and Electrical Engineering, University of Leeds, LS2 9JT, UK}
\affiliation{Institute of Physics, University of Belgrade, Pregrevica 118, 11080 Belgrade, Serbia}
\author{Dragan Indjin}
\affiliation{School of Electronic and Electrical Engineering, University of Leeds, LS2 9JT, UK}
\author{Vitomir Milanovi\'c}
\affiliation{School of Electrical Engineering, University of Belgrade, Bulevar kralja Aleksandra 73, 11120 Belgrade, Serbia}
\author{Jelena Radovanovi\'c}
\affiliation{School of Electrical Engineering, University of Belgrade, Bulevar kralja Aleksandra 73, 11120 Belgrade, Serbia}
\author{Zoran Ikoni\'c}
\affiliation{School of Electronic and Electrical Engineering, University of Leeds, LS2 9JT, UK}
\author{Paul Harrison}
\affiliation{School of Electronic and Electrical Engineering, University of Leeds, LS2 9JT, UK}



\date{\today}

\begin{abstract}
Magnetotunnelling spectroscopy of resonant tunnelling structures provides information on the nature of the two-dimensional electron gas in the well. We describe a model based on nonequilibrium Green's functions that allows for a comprehensive study of the density of states, tunnelling currents and current spin polarization. The investigated effects include the electron-phonon interaction, interface roughness scattering, Zeeman effect and the Rashba spin-orbit interaction. A qualitative agreement with experimental data is found regarding the satellite peaks. The spin polarization is predicted to be larger than ten percent for magnetic fields above 2 Tesla and having a structure even at the satellite peaks. The Rashba effect is confirmed to be observable as a beating pattern in the density of states but found to be too small to affect the tunnelling current.
\end{abstract}

%\keywords{}

\maketitle

\section{INTRODUCTION}

 Resonant tunnelling was first observed in scattering of electrons by noble gases where it is known as the Ramsauer effect\cite{bohm_quantum_theory_1989}.  After the invention of quantum mechanics, the appearance of the scattering minimum at a particular electron energy was explained by coupling with a quasi-bound state of the gas atom. Following the pioneering work of Esaki and Tsu\cite{tsu_book} in the seventies, the same effect was found in resonant tunnelling structures\cite{mizuta_rtd}(RTSs) where the role of electrostatic potential barriers is played by thin semiconductor layers.

 RTS is the basic nanoelectronic device that exploits the quantum nature of electrons. As the electron transport through a RTS is a combination of coherent tunnelling and phase-breaking collisions, RTSs are the ideal testing ground for quantum transport theories. In the envisioned high-performance electronic applications, one aims to maximize the main current peak where coherent transport dominates while minimizing the satellite peaks that occur due to scattering. Resonances appear in scattering as well because it is enhanced for the localized quasi-bound states, the most pronounced feature being the longitudinal optical (LO) phonon satellite peak\cite{goldman_lo-phonon_tunneling_rtd_1987}. While leading to performance degradation and power dissipation, the scattering mechanisms in RTSs are interesting in themselves for allowing the study of various aspects of the twodimensional electron gas (2DEG) physics via current spectroscopy. When a perpendicular magnetic field is applied, the 2DEG density of states (DOS) collapses into discrete Landau levels (LLs) thus pronouncing the resonant features in scattering. For this reason, magnetotunnelling experiments in RTSs \cite{mendez_magnetotunneling_1986,leadbeater_scattering_magnetic_field_rtd_1989,zaslavsky_magnetotunneling_double_barrier_haterostruct_1989,
  yang_magnetotunneling_spectroscopy_double_barr_diodes_1989,boebinger_observation_of_2d_magnetopolarons_in_resonant_tun_1990,
  zheng_nonresonant_magnetotunn_algaas_double_barr_1990,rascol_magnetospectral_analysis_double_barr_tunn_1990,chen_observation_of_2d_resonant_magnetopolarons_1991,goodings_phonon_assisted_magnetotunn_rtd_1993,wirner_phonon_effects_ingaas_rtd_magn_field_1997,
  popov_magnetotunneling_spectr_polarons_rtd_2010} have been established as an important tool in investigating the 2DEG physics.

  The electron-phonon interaction within the 2DEG of III-V RTSs leads to a formation of magnetopolarons\cite{das_sarma_theory_of_2d_magnetopolarons_1984} manifested as a characteristic anticrossing of LLs which has been observed in magnetotunnelling experiments\cite{leadbeater_scattering_magnetic_field_rtd_1989,zaslavsky_magnetotunneling_double_barrier_haterostruct_1989,
  yang_magnetotunneling_spectroscopy_double_barr_diodes_1989,boebinger_observation_of_2d_magnetopolarons_in_resonant_tun_1990,chen_observation_of_2d_resonant_magnetopolarons_1991}.
 Another important phenomenon in InAs-based RTSs is the spin-orbit interaction (SOI). Due to potential applications in semiconductor spintronics\cite{fabian_semiconductor_spintronics_2007}, the zero-magnetic field spin splitting in a 2DEG has been receiving considerable attention since the  field effect spin transistor was proposed by Datta and Das\cite{datta_das_transistor_1989}. At smaller magnetic fields SOI is known to significantly affect the 2DEG LLs leading to a beating pattern in the DOS. This causes a beating pattern in magnetoresistivity which is often used to characterize SOI in a 2DEG\cite{luo_observation_spin_splitting_gasb_inas_gasb_qw_1988,das_evidence_of_soi_ingaas_1989,engels_spin_splitting_in_ingaasinp_1997,grundler_2000}.

 Here we report on a theoretical study of magnetotunnelling  with LO phonon and interface roughness (IR) scattering taking a particular care of the carrier spin. The spin splitting is described as a combination of SOI and the Zeeman effect, both of which are known to be pronounced in InAs quantum wells. A detailed description is given of a nonequilibrium Green's function (NEGF) model\cite{mahan_quantum_transport_equation_1987,datta_kinetic_eq_1990, lake_negf_for_rtd_1992, datta_mesoscopic, lake_multiband_negf_1997,wacker_negf_vs_density_matrices_2001,lee_wacker_negf_for_qcls_2002,isic_jap_2010} with scattering treated within the self-consistent first Born approximation (FBA). The LO phonon and IR correlation functions are approximated with delta functions, the suitability of which is discussed in Appendices \ref{sec_IR_corr_funcs} and \ref{sec_LO_corr_funcs}. The model is applied to a symmetrical InAs-GaAs double barrier RTS and shown to yield I-V curves that qualitatively match the experimental data. An estimate of the spin polarization is found, and the role of SOI in spin-dependent transport through RTSs\cite{voskoboynikov_2000,koga_prl_2002,glazov_2005,isic_jap_2007} discussed.

\section{MODEL DESCRIPTION}

 In the first part we discuss the electronic states of the RTS in absence of scattering. Eigenstates of the lateral and spin degrees of freedom found here are used in the second part to formulate the transport model with scattering.

\subsection{Effective Hamiltonian and the Landau levels}

\begin{figure}
\includegraphics[width=10cm]{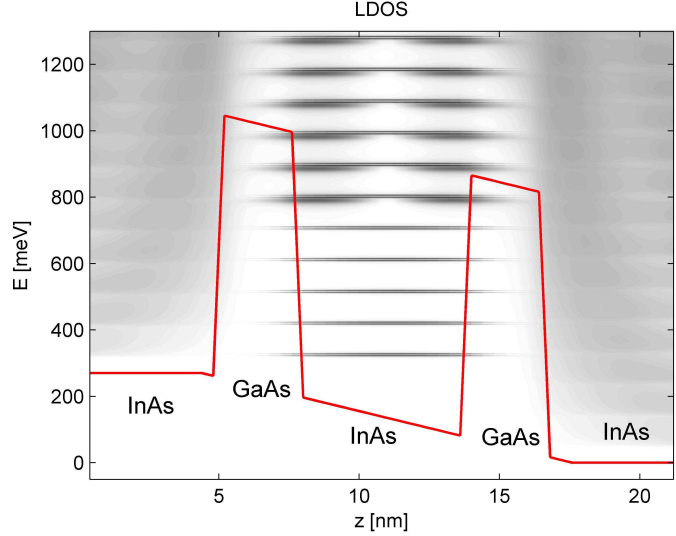}
\caption{\label{fig_rts_dos} The symmetric InAs-GaAs RTS under bias. The barrier and well widths are $3$ and $6\mathrm{nm}$, respectively. The conduction band edge with slope due to the external bias is shown with the full (red) line. The shading represents the LDOS in logarithmic scale. The pale series of peaks in the emitter (left) and collector (right) region represent Van Hove singularities at LL subband bottoms while the intense peaks in the well are the quasi-bound LLs. Up to around $700\mathrm{meV}$ only the lowest $z$-subband is present, the second $z$-subband states appearing as broadened double-node peaks.}
\end{figure}

The RTS is assumed to be grown along the $[100]$ direction denoted as the $z$-axis. It comprises of: (1) a highly doped InAs emitter contact, (2) the $3\mathrm{nm}$ wide left barrier formed by a GaAs layer, (3) the $6\mathrm{nm}$ InAs well, (4) the $3\mathrm{nm}$ wide right barrier made of GaAs and (5) a highly doped InAs collector contact. We adopt a one-band envelope function approximation (EFA) \cite{bastard_book} in which the electronic states of a heterostructure are determined by their effective mass $m^*$ and the effective potential $V(z)$ comprising the conduction band edge variation and the externally applied voltage. A generalized multiband EFA allows for many parameters and is applicable for modelling a wide range of semiconductor heterostructures \cite{voon_the_kp_method_book}. The one-band approximation used here is often adopted for single-valley conduction band heterostructure states. This is because up to few hundred milielectron volts above the conduction band minimum, the states are of a dominantly s-like character so both energy dispersion and matrix elements can be sufficiently well estimated using effective parameters. The use of more complicated models such as the multiband EFA or, especially with NEGF calculations, multiband empirical tight-binding models \cite{lake_multiband_negf_1997,klimeck_nemo3d_part1_2007} is justified when a more quantitative model\cite{bowen_quantitative_rtd_1997} is aimed for or even mandatory if effects like inter-valley scattering or inter-band tunnelling are studied.

 The conduction band effective potential $V(z)$ of the investigated RTS is depicted by the thick solid line in Fig. (\ref{fig_rts_dos}). The externally applied collector to emitter voltage $U_\mathrm{ce}$ is assumed to be positive so that electrons have a tendency to flow towards the collector whose band edge is taken as a reference energy. The EFA Hamiltonian of the biased RTS with an external magnetic field applied parallel to the growth direction reads
\begin{equation} \label{eq_hamiltonian}
H=\frac{\boldsymbol{\pi}^2}{2m^*}+V(z)+\frac{1}{2}g^*\mu_\mathrm{B}B\sigma_z+\frac{\alpha_R}{\hbar}(\pi_x\sigma_y-\pi_y\sigma_x).
\end{equation}
The first term is the kinetic energy, $\boldsymbol{\pi}=\mathbf{p}+e\mathbf{A}$ being the kinetic momentum of the electron and $m^*$ its conduction band effective mass. The third term represents the Zeeman effect determined by the effective $g$-factor, $g^*$, and $\sigma_{x,y,z}$ are the Pauli spin matrices \cite{bastard_book}.

 The fourth term is the Rashba spin-orbit interaction (RSOI) Hamiltonian that accounts for SOI due to the heterostructure inversion asymmetry  \cite{silva_1997}. The Dresselhaus SOI (DSOI) caused by the bulk inversion asymmetry is neglected. This is believed to hold in InAs quantum wells, considering the theoretical arguments in Ref. \onlinecite{lommer_spin_splitting_in_semicond_heterostruct_1988} and experimental findings in Refs. \onlinecite{luo_observation_spin_splitting_gasb_inas_gasb_qw_1988,luo_effects_of_ia_electrons_gasb_inas_gasb_qw_1990} (for InAs-GaSb quantum wells) and Refs. \onlinecite{nitta_gate_control_of_soi_1997,koga_rsoi_probed_by_waa_ingaalas_2002} (for InGaAs-InAlAs quantum wells). The RSOI in a heterostructure is a combined effect of the externally applied potential and the spatial variation of the band edge. In asymmetric heterostructures the latter lifts the spin degeneracy even in absence of an external electric field $E_\mathrm{ext}$. For symmetric structures the Rashba constant $\alpha_\mathrm{R}$ can be taken to be proportional to $E_\mathrm{ext}$
 \begin{equation} \label{eq_alpha_Cs}
 \alpha_\mathrm{R}=CE_\mathrm{ext}, \quad C=C_\mathrm{E}+C_\mathrm{I},
 \end{equation}
 where the direct contribution of the electric field is evaluated as  \cite{silva_1997}
 \begin{equation} \label{eq_Cs}
 C_\mathrm{E}=\frac{e\hbar^2}{2m^*}\frac{(2E_\mathrm{g}+\Delta_\mathrm{SO})\Delta_\mathrm{SO}}{E_\mathrm{g}(E_\mathrm{g}+\Delta_\mathrm{SO})(3E_\mathrm{g}+2\Delta_\mathrm{SO})}.
 \end{equation}
 The interface contribution $C_\mathrm{I}$ depends on details of the heterostructure potential. We have found that in the range of external fields of interest (around the main and satellite current peaks) taking into account only the first term gives $\alpha_\mathrm{R}$ in the range $1-2\approx 10^{-11}\mathrm{eVm}$ which is close to values for a InAlAs-InGaAs quantum well reported in \cite{grundler_2000} that vary in the range $2-5\approx 10^{-11}\mathrm{eVm}$. For this reason, unless it is varied as a parameter, $\alpha_\mathrm{R}$ is evaluated from Eqs. (\ref{eq_alpha_Cs}) and (\ref{eq_Cs}) with $E_\mathrm{g}(\mathrm{InAs})=356\mathrm{meV}$, $\Delta_\mathrm{SO}(\mathrm{InAs})=380\mathrm{meV}$ while $C_\mathrm{I}$ is set to zero.

 Thus, our model has three main effective parameters, $m^*$, $g^*$ and $\alpha_\mathrm{R}$. The energy scale over which these parameters change considerably is set by the bandgap. In InAs the effective mass increases approximately linearly from the zone center value $m^*=0.023m_0$ to around twice that at $E=0.5\mathrm{eV}$. The variation of the effective $g$-factor predicted by the $\mathbf{k}\cdot \boldsymbol{\pi}$ model is
   \begin{equation} \label{eq_g_fac_kp}
 g^*(E)=g^*(0)\frac{E_\mathrm{g}(E_\mathrm{g}+\Delta_\mathrm{SO})}{\Delta_\mathrm{SO}}\left(\frac{1}{E_\mathrm{g}+E}-\frac{1}{E_\mathrm{g}+\Delta_\mathrm{SO}+E}\right),
 \end{equation}
 with the zone center value $g^*(0)$ given by
 \begin{equation} \label{eq_g0}
 g^*(0)=\left(1-\frac{m_0}{m^*(0)}\right)\frac{2\Delta_\mathrm{SO}}{2\Delta_\mathrm{SO}+3E_\mathrm{g}}.
 \end{equation}
 $g^*(E)$ calculated from the above equations with InAs parameters is a monotonously increasing function starting at $g^*(0)=-16.9$ and going to around $-4$ at $0.5\mathrm{eV}$. The effective $g$-factor is also known to depend on the electron-electron interaction, thus a range of values has been reported in InAs quantum wells \cite{miura_high_magnetic_field_book_2008}. In Ref. \onlinecite{moller_spin_splitting_in_narrow_inas_qws_2003} $\vert g^* \vert$ of a 2DEG in a 4nm wide InAs quantum well has been found to be around 6. The situation with $\alpha_\mathrm{R}$ is less clear since it vanishes in bulk while a value in a heterostructure is highly dependent on both the constituents and electrostatic effects. It will be seen below that the mechanisms involved in electron transport through the investigated RTS have an energy scale of few tens of milielectron volts, determined by the chemical potential of the contacts and the InAs LO phonon energy.  Consequently, keeping in mind the range of meaningful values, it makes sense to fix the effective parameters which is done by setting $m^*(\mathrm{InAs})=0.023m_0$, $g^*=-5$ and calculate $\alpha_\mathrm{R}$ as explained above.

 \begin{figure}
\includegraphics[width=15cm]{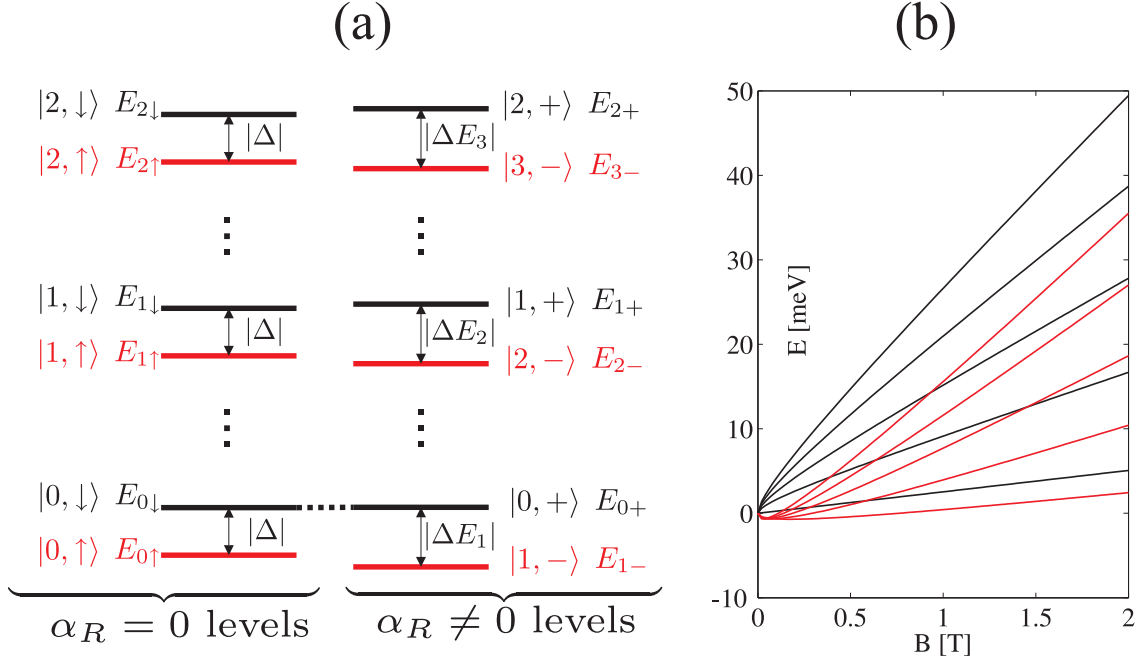}
\caption{\label{fig_energy_levels} A scheme of the $E_{ns}$ levels and their relation with the ordinary ($\alpha_\mathrm{R}=0$) LLs for negative $g$. (a) Shows the arrangement of the levels at high magnetic fieds. (b) is the dispersion $E_{ns}(B)$ for small and moderate magnetic fields. For clarity, (b) is drawn with exaggerated $\alpha_\mathrm{R}$. }
\end{figure}

 In absence of RSOI ($\alpha_\mathrm{R}=0$), the eigenkets describing the lateral $\boldsymbol{\rho}=(x,y)$ and spin degrees of freedom are the spin split LLs $\vert  n\sigma\rangle$ with energies
 \begin{equation} \label{eq_LL0_ens}
 E_{n\sigma}=\left(n+\frac{1}{2}\right)\hbar \omega_\mathrm{C}\pm \frac{\Delta_B}{2}, \quad n=0,1,..., \quad \Delta_B=g^*\mu_B B, \quad \omega_\mathrm{C}=\frac{eB}{m^*},
 \end{equation}
 where the plus is to be taken for spin up, $\sigma=\uparrow$, and the minus for spin down along the $z$ axis, $\sigma=\downarrow$. Adopting the Landau gauge, the corresponding wavefunctions are
 \begin{equation} \label{eq_LL0_psis}
 \langle \boldsymbol{\rho} \uparrow \vert kn \uparrow \rangle =\langle \boldsymbol{\rho} \downarrow \vert kn \downarrow \rangle=\frac{e^{ikx}}{\sqrt{L_x}}\Phi_n\left(\frac{y}{L}-kL\right), \quad
  \langle \boldsymbol{\rho} \uparrow \vert kn \downarrow \rangle=\langle \boldsymbol{\rho} \downarrow \vert kn \uparrow \rangle =0.
 \end{equation}
 In this equation $k$ and $L_x$ denote the quasi-momentum and system length along the $x$ axis, while $L=\sqrt{\hbar/eB}$ is the magnetic length. The quantization of the lateral degrees of freedom is manifested in the local density of states (LDOS) shown in Fig. \ref{fig_rts_dos} where it leads to a series of Van Hove singularities (corresponding to the bottom of each LL subband) in the emitter and collector and to a series of quasi-discrete levels in the well.

 For nonzero RSOI the LLs given by Eqs. (\ref{eq_LL0_ens}) and (\ref{eq_LL0_psis}) are mixed into eigenstates labelled as $\vert kns\rangle$ where the allowed values of the pair of quantum numbers $n$ and $s$ are \cite{schliman_quantum_hall_and_soi_2003,wang_2deg_magnetotransp_soi_2005,yang_interplay_between_soi_and_zeeman_splitting_2006}
 \begin{equation}
 ns=0+,1-,1+,2-,2+,....
 \end{equation}
 The corresponding energies are
 \begin{equation} \label{eq_ens}
E_{ns}=n\hbar\omega_\mathrm{C}+s\sqrt{\left(\frac{\hbar\omega_\mathrm{C}-\Delta_B}{2}\right)^2+nR^2}, \quad R=\sqrt{2}\alpha_\mathrm{R}/L.
\end{equation}
  The $\vert kns\rangle$ states can be expressed in terms of $\vert kn\sigma\rangle$ as
 \begin{equation} \label{eq_kns_kets}
 \vert kns \rangle = A^\uparrow_{ns}\vert k(n-1)\uparrow \rangle+A^\downarrow_{ns}\vert kn\downarrow\rangle,
 \end{equation}
 with coefficients $A^\sigma_{ns}$ given by
 \[
A^\uparrow_{n+}=\frac{\sqrt{n}R}{\sqrt{nR^2+(E_{n+}-E_{n-1\uparrow})^2}}, \quad
A^\downarrow_{n+}=\frac{i(E_{n+}-E_{n-1\uparrow})}{\sqrt{nR^2+(E_{n+}-E_{n-1\uparrow})^2}},
\]
\begin{equation}
A^\uparrow_{n-}=\frac{i(E_{n\downarrow}-E_{n-})}{\sqrt{nR^2+(E_{n\downarrow}-E_{n-})^2}}, \quad
A^\downarrow_{n-}=\frac{\sqrt{n}R}{\sqrt{nR^2+(E_{n\downarrow}-E_{n-})^2}}.
\end{equation}
In evaluating Eq. (\ref{eq_kns_kets}) for for $ns=0+$, $\vert k(n-1)\uparrow \rangle$ is to be taken as the zero ket. The arrangement of the $E_{n\sigma}$ and $E_{ns}$ energy levels for a negative effective $g$-factor is depicted in Fig. \ref{fig_energy_levels}. By analyzing Eqs. (\ref{eq_ens}) and (\ref{eq_kns_kets}), several observations can be made regarding the effect of RSOI on LLs : (1) the $\vert 0+\rangle$ ket is identical (apart from the irrelevant phase factor) to $\vert 0\downarrow\rangle$ and they have the same energy, $E_{0+}=E_{0\uparrow}$, as depicted by the dots connecting these two levels in Fig. \ref{fig_energy_levels}; (2) the two closely spaced spin-split levels $E_{n\uparrow}$ and $E_{n\downarrow}$ become $E_{(n+1)-}$ and $E_{n+}$, respectively; (3) the Zeeman spin splitting, $\Delta_B$, becomes $\Delta E_n= E_{n+}-E_{(n+1)-}$ which increases with $n$ and is greater than $\Delta_B$ for all $n$.

\subsection{The quantum transport model}

  Initially, magnetotunnelling in RTS has been studied \cite{silva_resonant_tunneling_via_LLs_1988,schulz_resonant_tunneling_through_LLs_inelastic_scatt_broadening_1990,wang_sequential_approach_to_magnetotunneling_1994} within the frame of the Landauer-B\"{u}ttiker formalism neglecting the phase-breaking scattering altogether or using ad hoc schemes to describe it \cite{datta_mesoscopic}. In Refs. \onlinecite{zou_phonon_magnetotunneling_1993,bo_finite_temp_magnetotunneling_1995,bo_phonon-assisted_magnetotunneling_negf_1997} the Matsubara technique suitable for a rigorous treatment of many-body effects at finite temperatures\cite{bruus_many-body} is used to evaluate the RTS transmission probability in the presence of LO phonon scattering. The combined electrostatic charging and LO phonon scattering effects have been studied in Ref. \onlinecite{orellana_self-const_calc_magnetotunn_1996} by evaluating a conveniently devised equation of motion for the electron field operators. While the degree of complexity and ability to account for various physical phenomena of the models used in these studies vary, their common difficulty is a lack of a consistent description of the non-equilibrium state of the biased RTS. It is for this reason that the NEGF model that simultaneously accounts for the quantum mechanical details and  the non-equilibrium statistics is the principal method in studying the RTS. Existing NEGF studies of RTS magnetotunnelling are scarce and include the study of scattering on LO phonons \cite{zhou_scattering_magnetotunneling_qw_1996} and the effect of shallow impurity states \cite{gutierrez_magnetotunneling_through_shallow_impurity_states_double_barr_2004}.

 The quantities used in the NEGF method are the Green's functions $G^\gamma$ and self-energies $\Sigma^\gamma$ both of which can be of three kinds: lesser ($\gamma=<$), greater ($\gamma=>$) and retarded ($\gamma=R$). In explaining the model we need to jump from the mixed, $\vert kns\rangle\vert z\rangle$, to the real-space, $\vert \boldsymbol{\rho}\sigma\rangle \vert z\rangle=\vert\mathbf{r}\sigma\rangle$, representation and vice versa, so to improve the paper readability we start by giving a step-by-step explanation of how the equation for determining $G^R$ is arrived at.

 The retarded Green's function $G^R$ is defined in the time domain as\cite{bruus_many-body}
 \begin{equation}
 G^R_{s_1s_2}(k_1n_1z_1,k_2n_2z_2;t_1,t_2)=-\frac{i}{\hbar}\theta(t_1-t_2)\left\langle \left\{\Psi_{s_1}(k_1n_1z_1;t_1),\Psi^\dag(k_2n_2z_2;t_2)\right\}\right\rangle.
 \end{equation}
 Here $t_{1,2}$ are the time coordinates, $\theta(t)$ is the Heaviside step function, $\langle ... \rangle$ is the ensemble average and $\left\{ ... \right\}$ the fermion anticommutator. $\Psi_{s}(knz,t)$ and $\Psi^\dag_{s}(knz,t)$ are the destruction and creation operators for the $\vert kns\rangle$ state at coordinate $z$ and time $t$. The spin coordinate ($s=\pm$) is arbitrarily written as a subscript. As the problem under consideration is stationary, all the quantities depend only on the time difference $t_1-t_2$ and we switch to the energy domain, e.g.
 \begin{equation} \label{eq_time_to_e}
 G^R_{s_1s_2}(k_1n_1z_1,k_2n_2z_2;E)=\int d(t_1-t_2)e^{iE(t_1-t_2)/\hbar}G^R_{s_1s_2}(k_1n_1z_1,k_2n_2z_2;t_1-t_2).
 \end{equation}
 Since the Hamiltonian (\ref{eq_hamiltonian}) allows for decoupling the lateral and the $z$ degrees of freedom while $\vert kns\rangle$ are eigenstates, $G^R$ is diagonal in $kns$ in absence of scattering (any perturbation to $H$). The onset of a scattering mechanism generally mixes the $\vert kns\rangle$ states so off-diagonal terms may appear. However, depending on the scattering model details, the coupling between any given $\vert k_1n_1s_1\rangle$ and all the other $\vert kns\rangle$ states induced by the perturbation can often be properly described by a diagonal term of the self-energy $\Sigma^R$. All the Green's functions and self-energies considered here are diagonal in $k$ and $n$ but not in $s$, so the full notation from Eq. (\ref{eq_time_to_e}) will be abbreviated by $G^R_{s_1s_2}(knz_1z_2;E)$. This (quasi-) decoupling of the $kn$ coordinates will be clarified later after the scatterer correlation functions are introduced.

 The retarded Green's function is found by fixing $kn$ and inverting the matrix
 \begin{equation} \label{eq_gr}
 G^R_{s_1s_2}(knz_1z_2;E)=\left(E-H_{s_1s_2}(knz_1z_2)-\Sigma^R_{s_1s_2}(knz_1z_2;E)\right)^{-1},
 \end{equation}
 where the $z$ degree of freedom is represented by a tight-binding model with the nearest-neighbour hopping energy\cite{datta_atom} $t=\hbar^2/2m^*a^2$ and care taken to ensure Hermiticity around the heterointerfaces where the effective mass changes abruptly\cite{lake_multiband_negf_1997}. It is equivalent to replacing the derivatives along $z$ by finite-differences defined on a set of equidistant points with $a$ being the point-to-point distance. Therefore, if the $z$ axis interval enclosing the RTS is represented by $N$ mesh points, Eq. (\ref{eq_gr}) is a $2N\times 2N$ matrix equation, the doubling being due to spin.

 The self-energies in (\ref{eq_gr}) are a sum of four terms that are calculated self-consistently
 \begin{equation}
 \Sigma^\gamma=\Sigma^\gamma_\mathrm{E}+\Sigma^\gamma_\mathrm{C}+\Sigma^\gamma_\mathrm{IR}+\Sigma^\gamma_\mathrm{LO}.
 \end{equation}
 The emitter ($\Sigma^\gamma_\mathrm{E}$) and collector ($\Sigma^\gamma_\mathrm{C}$) self-energies describe the interaction of electrons within the RTS with the exterior of the device, as explained in Ref. \onlinecite{datta_mesoscopic}. The interface roughness ($\Sigma^\gamma_\mathrm{IR}$) and LO phonon ($\Sigma^\gamma_\mathrm{IR}$) self-energies are assumed zero in the first iteration.

 The lesser ($\lambda=<$) and greater ($\lambda=>$) Green's functions are found from the kinetic equation
 \begin{equation}
 G^\lambda_{s_1s_2}(knz_1z_2;E)=\sum_{s_3,s_4}\int_{z_\mathrm{E}}^{z_\mathrm{C}} dz_3 \int_{z_\mathrm{E}}^{z_\mathrm{C}}dz_4 G^R_{s_1s_3}(knz_1z_3;E)\Sigma^\lambda_{s_3s_4}(knz_3z_4;E)
 G^A_{s_4s_2}(knz_4z_2;E),
 \end{equation}
 where $G^A$ is the advanced Green's function (in the energy domain, it is the complex conjugate of $G^R$). In subsequent iterations FBA is used to evaluate the interface roughness ($\phi=\mathrm{IR}$) and phonon ($\phi=\mathrm{LO}$) lesser and greater self-energies
 \begin{equation} \label{eq_fba}
 \Sigma^\lambda_{\phi,\sigma\sigma}(\mathbf{r}_1\mathbf{r}_2;E)=\frac{1}{2\pi}\int \frac{dE'}{\hbar} G^\lambda_{\sigma\sigma}(\mathbf{r}_1\mathbf{r}_2;E-E')D^\lambda_{\phi}(\mathbf{r}_1\mathbf{r}_2;E'),
 \end{equation}
 where $D^\lambda_{\phi}(\mathbf{r}_1\mathbf{r}_2;E)$ denotes the scatterer correlation function assumed to be independent on the electron spin (we neglect the scattering via SOI). For IR scattering the lesser ($D^<$) and greater ($D^>$) correlation functions are equal
 \begin{equation}\label{eq_local_DIR}
 D_{\mathrm{IR}}(\mathbf{r}_1\mathbf{r}_2;E)=2\pi\hbar \delta(E)\delta(\mathbf{r}_1-\mathbf{r}_2)\sum_{z_I} U_\mathrm{IR}^2(z_\mathrm{I})\delta(z-z_\mathrm{I}),
 \end{equation}
 where the summation is done over the heterointerfaces located at $z=z_\mathrm{I}$. The LO-phonon correlation functions are taken to be
  \begin{equation} \label{eq_local_DLO}
 D_\mathrm{LO}^\gtrless(\mathbf{r}_1,\mathbf{r}_2;\omega)=2\pi\hbar \delta(E\mp E_\mathrm{LO}) U_\mathrm{LO}^2 \delta(\mathbf{r}_1-\mathbf{r}_2),
\end{equation}
 where $E_\mathrm{LO}=29\mathrm{meV}$ is the LO phonon energy in InAs. The delta-like form of correlation functions given by Eqs. (\ref{eq_local_DIR}) and (\ref{eq_local_DLO}) is used because it considerably simplifies the calculations\cite{lake_negf_for_rtd_1992,lake_phonon_peak_1993} by decoupling the equations for states with different $kn$ coordinates. To clarify the approximation being made, note that the IR correlation function used in the literature is usually a Gaussian function\cite{ando_2deg_1982} with the deviation set by the roughness lateral length $\Lambda$ (typically few nanometers), while the Fr\"{o}hlich interaction with LO phonons gives a dependence proportional to\cite{mahan_1990} $1/\vert \mathbf{r}_1-\mathbf{r}_2\vert$. Roughly, the approximation consists in assuming that the self-energy at a given point in space is determined solely by the electronic properties at that same point while the more realistic models allow for a contribution from all the nearby points. A quantitative discussion of scatterer correlation functions aimed at determining meaningful values for scattering strengths $U_\mathrm{IR}$ and $U_\mathrm{LO}$ is given in Appendices \ref{sec_IR_corr_funcs} and \ref{sec_LO_corr_funcs}.

 The mixed representation of the lesser and greater self-energies is then found as
\begin{equation} \label{eq_sigmaIR_lambda}
\Sigma_{\mathrm{IR}s_1s_2}^\gtrless(kn,z_1,z_2;\omega)=\delta(z_1-z_2)\sum_\mathrm{I}U_\mathrm{IR}^2(z_\mathrm{I})\delta(z_1-z_I)\sum_\sigma G^\gtrless_{\sigma\sigma}({\boldsymbol\rho}_1z_1,{\boldsymbol\rho}_1z_2;E)A^{\sigma*}_{ns_1}A_{ns_2}^{\sigma},
\end{equation}
for IR scattering and
\begin{equation} \label{eq_sigmaLO_lambda}
\Sigma_{\mathrm{LO}s_1s_2}^\gtrless(n,z_1,z_2;\omega)=\delta(z_1-z_2)U_\mathrm{LO}^2\sum_\sigma G^\gtrless_{\sigma\sigma}({\boldsymbol\rho}_1z_1,{\boldsymbol\rho}_1z_2;E\mp E_\mathrm{LO})A^{\sigma*}_{ns_1}A_{ns_2}^{\sigma},
\end{equation}
for LO phonons. The real-space representation of Green's functions in the above equations is given by
 \begin{equation} \label{eq_GF_realspace}
 G_{\sigma\sigma}^\gamma({\boldsymbol\rho}_1z_1,{\boldsymbol\rho}_1z_2;\omega)=\frac{1}{2\pi L^2}\sum_{ns_1s_2} A^\sigma_{ns_1}A^{\sigma *}_{ns_2} G^\gamma_{s_1s_2}(n,z_1,z_2;\omega), \quad \sigma=\uparrow,\downarrow.
 \end{equation}

 The retarded self-energy is related to the lesser and greater self-energies by
 \begin{equation} \label{eq_retarded_self_energy}
 \Sigma^R(E)=\frac{1}{2}\left[\Sigma^>(E)-\Sigma^<(E)\right]+\frac{i}{2\pi}\mathcal{PV}\int dE'\frac{\Sigma^>(E')-\Sigma^<(E')}{E-E'},
 \end{equation}
 where $\mathcal{PV}$ denotes the Cauchy principal value. The left term is the anti-Hermitian part that causes the broadening of energy levels. The right term is Hermitian and corresponds to the scattering-induced energy shift (the energy correction found by time-independent perturbation theory \cite{sakurai_modern}). In studying the electron transport through a RTS, the energy shift (of the order of a milielectron volt) due to various interactions is not very significant especially since the exact quasi-bound state energies are hard to estimate anyway. On the other hand, the level broadening leads to qualitative differences and its exact value is crucial in estimating the degree of spin polarization. Therefore, in order to avoid the difficulties in the numerical evaluation of the Cauchy principal value, the Hermitian part is neglected, as often done in NEGF calculations\cite{vukmirovic_negf_prb_2007}.

 Once the values of $\Sigma^R$ are found, the calculations are repeated to obtain a self-consistent solution, as in Ref. \onlinecite{isic_jap_2010}.

  \section{RESULTS AND DISCUSSION}

  Magnetotunnelling through a RTS is governed by the coupling of the emitter and collector electron baths via the discrete states in the well, the coupling being a combined effect of coherent tunnelling through the barriers and incoherent transport assisted by scattering on the interface roughness and LO phonons. Consequently, the starting point of our discussion is the (quasi)discrete LL spectrum in the well, which is described by the spin-polarized DOS in the RTS. After clarifying the structure of this spectrum, we move on to study the properties of the vertical transport and discuss the features in the RTS I-V curves. Throughout this section low temperature ($T=4.2\mathrm{K}$) with equal emitter and collector electrochemical potentials ($\mu_\mathrm{E}=\mu_\mathrm{C}=20\mathrm{meV}$) are assumed.

 \subsection{Properties of the RTS DOS}

  In absence of RSOI, the dispersion of LLs, $E_{n\sigma}(B)$, is represented by the usual Landau fan diagram. The presence of RSOI leads to a nonlinear dispersion $E_{ns}(B)$ given by Eq. (\ref{eq_ens}) and plotted for first few LLs in Fig. \ref{fig_energy_levels}. By investigating Eq. (\ref{eq_ens}) it is seen that the arrangement of $E_\mathrm{ns}$ levels shown schematically in Fig. \ref{fig_energy_levels} is reached at $B$ large enough that $\hbar \omega_\mathrm{C} \gg \sqrt{n} R$. The large spacing between the doublet $E_{n+}$ and $E_{(n-1)-}$ and any other level implies a simple DOS structure consisting of a series of double peaks (separated by the spin-splitting energy $\Delta E_n=E_{n+}-E_{(n-1)-}$) arranged at a distance $\hbar \omega_\mathrm{C}$ apart. However, a rich structure is seen in DOS at smaller magnetic fields.

  Considering that $\hbar \omega_\mathrm{C}$ is proportional to $B$ and that $R$ is proportional to $\sqrt{B}$, at $B\rightarrow 0$ all the $E_{n-}$ levels lie below all the $E_{n+}$ levels ($E_{0+}$ being the lowest amongst them). With increasing $B$, each $E_{n-}$ level approaches the corresponding $E_{(n+1)+}$ level crossing all the $E_{n+}$ levels below it. This produces a characteristic beating pattern in the DOS for any fixed $B$\cite{wang_2deg_magnetotransp_soi_2005}. While the exact energy spacing $\Delta E$ between the two lowest nodes is easily calculated from the $E_{ns}(B)$ dispersion, to understand its order of magnitude and dependence on $B$ and the RSOI parameter $\alpha_\mathrm{R}$ we note that the crossing occurs when the energy difference $\hbar \omega_\mathrm{C}$ between two successive LLs is compensated by $\sqrt{n} R$. Therefore $n\sim (\hbar \omega_\mathrm{C}/R)^2$ and
  \begin{equation} \label{eq_deltaE_sim}
  \Delta E \sim n \hbar\omega_\mathrm{C}=(\hbar \omega_\mathrm{C})^3/R^2,
  \end{equation}
  where the tilde denotes the order of magnitude. As an example, for $\alpha_\mathrm{R}=0.936\times10^{-11}\mathrm{eVm}$, $B=0.5\mathrm{T}$ (giving $\hbar\omega_\mathrm{C}\approx 2.39\mathrm{meV}$), Eq. (\ref{eq_deltaE_sim}) predicts $\Delta E\sim 99.7\mathrm{meV}$.

    The spin-polarized DOS in the RTS is obtained by integrating the spin-polarized LDOS over the length of the RTS
  \begin{equation}
  \mathrm{DOS}(E)_\sigma=\int_{z_\mathrm{E}}^{z_\mathrm{C}} dz \mathrm{LDOS}_\sigma(E,z),
  \end{equation}
  with
  \begin{equation}
  \mathrm{LDOS}_\sigma(E,z)=\frac{i}{2\pi}\left(G^R_{\sigma\sigma}(\boldsymbol{\rho}z,\boldsymbol{\rho}z;E)-G^A_{\sigma\sigma}(\boldsymbol{\rho}z,\boldsymbol{\rho}z;E)\right),
  \end{equation}
  and the total DOS is a sum of its spin polarized components, $\mathrm{DOS}(E)=\mathrm{DOS}_\uparrow(E)+\mathrm{DOS}_\downarrow(E)$.

   \begin{figure}
\includegraphics[width=16cm]{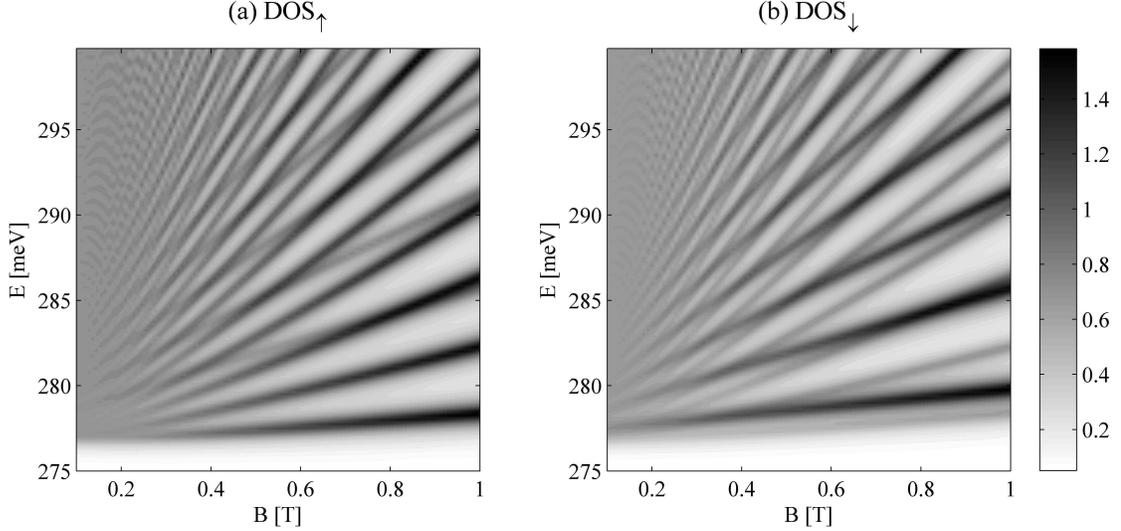}
\caption{\label{fig_LL_crossing_bal} Spin-polarized DOS of the RTS biased with $U_\mathrm{ce}=270\mathrm{mV}$ showing the dispersion $E_{ns}(B)$. To observe the LLs in this range of magnetic fields, the LLs must be very narrow. Here the ballistic limit is taken while the value of $\alpha_\mathrm{R}$ is exaggerated to five times the value predicted by Eq. \ref{eq_Cs} (approximately $4.6\times 10^{-11}\mathrm{eVm}$). The plotted quantity is $\log\left[1+DOS_\sigma(E)/D_{2\mathrm{DEG}}\right]$.}
\end{figure}

  There are two conditions that need to be satisfied for observing the crossing of the few lowest LLs: (1) $R\sim \hbar \omega_\mathrm{C}$ and (2) $ \hbar \omega_\mathrm{C}>\Gamma$, where $\Gamma$ denotes the LL broadening. The latter condition ensures that $B$ is large enough for a discrete spectrum to appear. In a RTS, $\Gamma$ is nonzero even for noninteracting electrons due to the finite probability of electron escaping from the well via tunnelling through barriers. For the structure shown in Fig. \ref{fig_rts_dos}, $\Gamma=0.8\mathrm{meV}$ in the ballistic limit (scattering neglected). Therefore, the LLs become clearly resolved for fields around $0.5\mathrm{T}$ ($\hbar\omega_\mathrm{C}\approx 2.5\mathrm{meV}$). At external voltages around the main current peak ($U_\mathrm{ce}=270\mathrm{mV}$), from Eq. (\ref{eq_Cs}) we find $\alpha_\mathrm{R}=10^{-11}\mathrm{eVm}$, giving $R\approx 0.4\mathrm{meV}$, so the crossing would not be observable. For purposes of illustration, Fig. \ref{fig_LL_crossing_bal} shows the spin-polarized DOS of the RTS for external bias $U_\mathrm{ce}=270\mathrm{mV}$ (main peak of the current) calculated in the ballistic limit and with $\alpha_\mathrm{R}$ set to five times the value given by Eq. (\ref{eq_Cs}) so that the crossing of LLs is clearly seen. The 2DEG spin-polarized DOS at zero magnetic field is given by $D_\mathrm{2DEG}=m^*/2\pi\hbar^2$. The energy is always measured relative to the collector band edge, so that the emitter band edge is $U_\mathrm{e}=eU_\mathrm{ce}$. If the RTS had a DOS as shown in Fig. \ref{fig_LL_crossing_bal} it would mean that the RSOI could be investigated by studying the structure of the RTS I-V curves. However, considering that the realistic value of $\Gamma$ is few meV and that the effective mass in an InAs quantum well is likely to be higher than $0.023m_0$, we conclude that it is not realistic to expect that the RSOI effects would affect the shape of the I-V curves.

 \begin{figure}
\includegraphics[width=10cm]{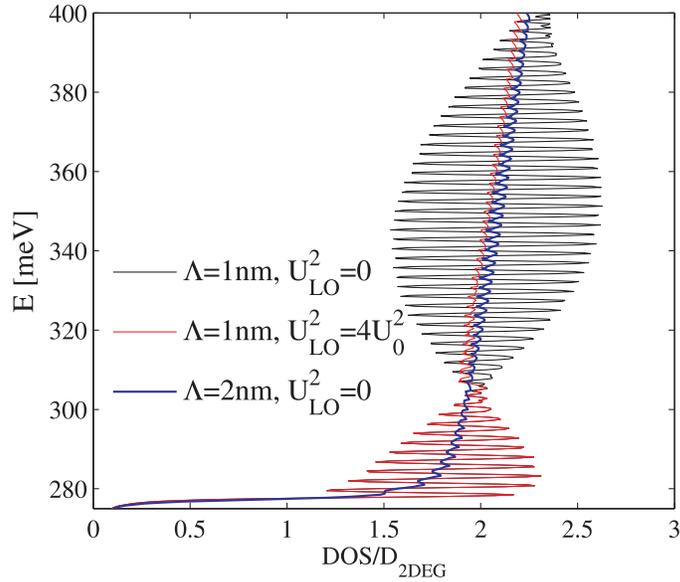}
\caption{\label{fig_DOS_beating} DOS of RTS biased with $U_\mathrm{ce}=270\mathrm{mV}$ and $B=0.5\mathrm{T}$. The red and black curves were calculated for a moderate IR scattering intensity ($\Lambda=1\mathrm{nm}$). The difference between the two is the onset of LO phonon induced broadening at energies more than $E_\mathrm{LO}$ above the emitter conduction band bottom. In SdH measurements this effect is not observed since the LO phonon emission at the Fermi level is blocked by the Pauli exclusion principle. The blue curve is the case with higher IR scattering, showing that the beating pattern exists even when the LL broadening is larger than $\hbar \omega_\mathrm{C}$. }
\end{figure}

 When scattering is included with strengths $U_\mathrm{IR}^2$ and $U_\mathrm{LO}^2$ as found in Appendices \ref{sec_IR_corr_funcs} and \ref{sec_LO_corr_funcs}, the broadening for magnetic fields around $1\mathrm{T}$ becomes $\Gamma \approx 3.5\mathrm{meV}$. It causes the overlapping of individual LLs for smaller magnetic fields and, consequently, makes the observation of the nonlinear features of the $E_{ns}$ dispersion impossible. However, RSOI can be experimentally observed via the beating pattern in Shubnikov-de Haas (SdH) oscillations\cite{luo_observation_spin_splitting_gasb_inas_gasb_qw_1988,das_evidence_of_soi_ingaas_1989,engels_spin_splitting_in_ingaasinp_1997,grundler_2000}. It is a consequence of the beating pattern seen in the DOS. This is illustrated in Fig. \ref{fig_DOS_beating}.  The beating pattern survives even with significant scattering effects. The distance between the nodes is found to be around $100\mathrm{meV}$, in good agreement with the estimate given by Eq. (\ref{eq_deltaE_sim}).

\begin{figure}
\includegraphics[width=15cm]{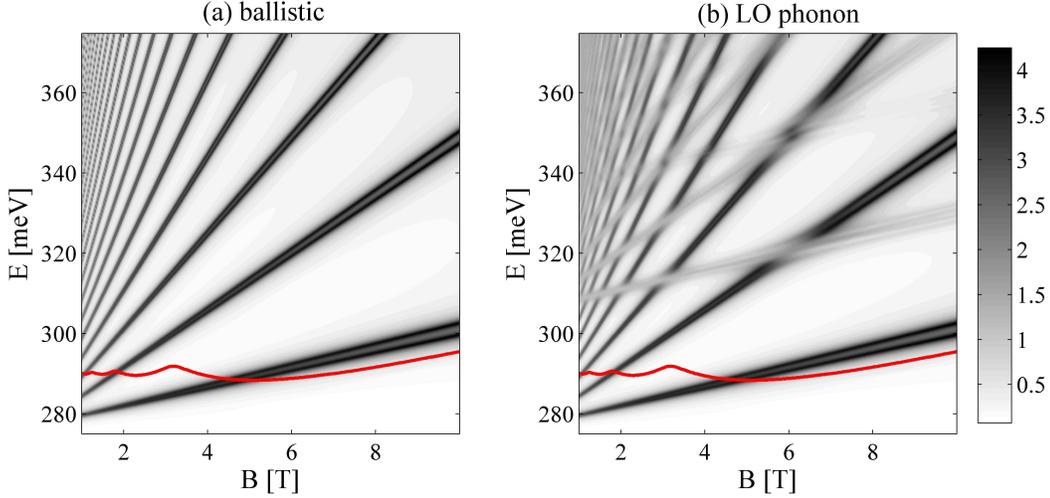}
\caption{\label{fig_DOS_phonon} The evolution of DOS with $B$ for $B>1$. The RTS is biased with $U_\mathrm{ce}=270\mathrm{mV}$. (a) Ballistic limit where the LL broadening is entirely due to coherent coupling with emitter and collector states. (b) DOS when the LO phonon scattering is included with $U_\mathrm{LO}^2=4U_0^2$. The red line shows the emmiter Fermi level. The Pauli blocking of the first replica of the lowest LL level below $4\mathrm{T}$ (when it is below the red line) is not complete because the system is not in equilibrium so the lowest LL is only partially occupied. The plotted quantity is $\log\left[1+DOS(E)/D_{2\mathrm{DEG}}\right]$.}
\end{figure}

 The effect of LO-phonon scattering becomes apparent if the DOS is observed on a larger scale. Figure \ref{fig_DOS_phonon} shows the (a) ballistic and (b) LO-phonon scattering DOS of the biased RTS. Compared to Fig. \ref{fig_DOS_phonon} (a), additional energy levels are seen in (b). These are the phonon replicas of the LLs displaced from them by multiples of $E_\mathrm{LO}$. The problem of electron-LO phonon interaction in a 2DEG, being of fundamental significance, has received a considerable attention\cite{das_sarma_theory_of_2d_magnetopolarons_1984,larsen_correction_energy_2d_magnetopolarons_1984,peeters_2d_and_3d_magnetopolarons_1985}. Polaron effects have been studied using cyclotron resonance \cite{sigg_polaron_effects_algaas_2deg_cycl_res} in a quantum well, or by investigating the I-V curve features in magnetotunnelling in perpendicular \cite{boebinger_observation_of_2d_magnetopolarons_in_resonant_tun_1990,chen_observation_of_2d_resonant_magnetopolarons_1991} and in-plane\cite{popov_magnetotunneling_spectr_polarons_rtd_2010} magnetic fields.

 The DOS shown in Fig. \ref{fig_DOS_phonon} (b) demonstrates that the current model captures some of the main features of the electron (magnetopolaron) DOS in a (quasi)2DEG: (1) appearance of the phonon replicas, (2) the anticrossing of LLs and their replicas at magnetophonon resonances and (3) a considerable broadening of LLs only around the magnetophonon resonance. Effects related to the real part of the self-energy, see Eq. (\ref{eq_retarded_self_energy}), such as the polaron shift and effective mass renormalization, are not included in this model. The LO phonon scattering strength $U_\mathrm{LO}^2$ determines the height of the DOS corresponding to phonon replicas. It also affects the anticrossing energy. A quantitative connection between the LO-phonon self-energy and $U_\mathrm{LO}^2$ is not simple to establish, owing to the fact that the self-energy is found as a self-consistent solution, but a rough estimate is that the self-energy scales as the square root of $U_\mathrm{LO}^2$. The significance of the phonon replicas in the RTS is that they provide an additional channel for electrons tunnelling through the RTS. For example, the lowest spin-split replicas seen in Fig. \ref{fig_DOS_phonon} (b) correspond to the lowest lying $\vert 0+\rangle$ and $\vert 1-\rangle$ LLs. Once their energies are aligned with electrons in the emitter having the same lateral coordinates, $ns=0+$ and $ns=1-$, resonant intra-LL (preserving the LL coordinates) tunnelling with LO phonon emission occurs.

  \begin{figure}
\includegraphics[width=15cm]{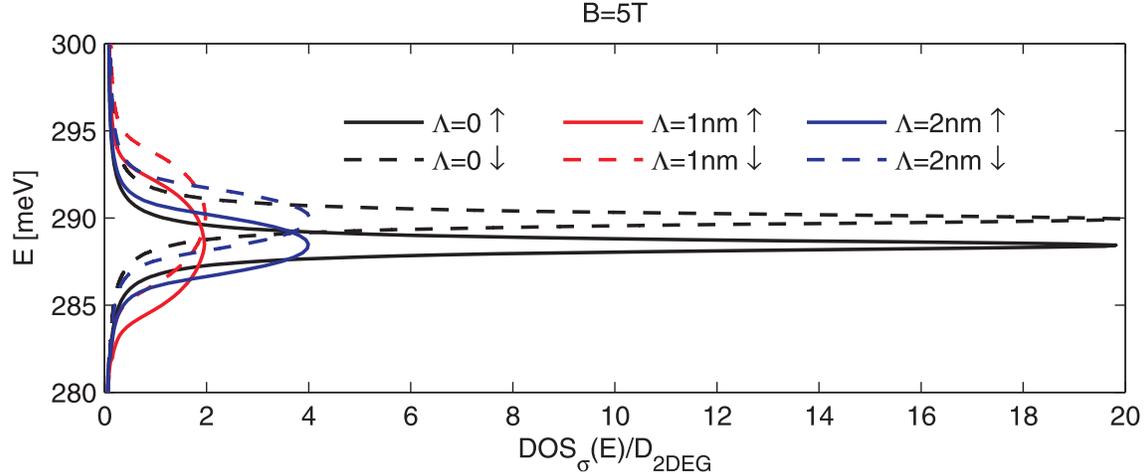}
\caption{\label{fig_dos_ir_effects} The effect of IR on spin-polarized DOS, $\mathrm{DOS}_\uparrow$ is shown in full lines and $\mathrm{DOS}_\downarrow$ in dashed lines. The bias voltage is $U_\mathrm{ce}=270\mathrm{mV}$, while the magnetic field is $B=5\mathrm{T}$. }
\end{figure}

 The effect of IR scattering on the DOS is simpler and consists of broadening all the states (both the LLs and their replicas). In contrast to the phonon scattering, it is known to depend only on the DOS and not on the occupation of the levels\cite{lake_multiband_negf_1997}. In Appendix \ref{sec_IR_corr_funcs} we argue that if a delta model is used for the roughness correlation, the scattering strength at interface $z_\mathrm{I}$ should be taken as
 \begin{equation}
 U_\mathrm{IR}=V_\mathrm{b}^2\Delta_\mathrm{IR}^2\pi \Lambda^2.
 \end{equation}
 In this approach only the product of $\Delta_\mathrm{IR}$ and $\Lambda$ is significant, so we fix $\Delta_\mathrm{IR}=0.3\mathrm{nm}$ and consider the effect of changing $\Lambda$. Values of $\Lambda=5-7\mathrm{nm}$ have been reported for AlGaAs quantum wells\cite{sakaki_interface_roughness_scatt_in_algaas_qw_1987,gottinger_interface_roughness_gaas_qws_1988} and yield a broadening of more than $15\mathrm{meV}$. Since the typical level broadening in a high quality RTS is expected to be several\cite{mizuta_rtd} meV, smaller values of $\Lambda$ have been used in calculations. In Fig. \ref{fig_dos_ir_effects} a comparison is made of the spin-polarized DOS in the biased RTS at $B=5\mathrm{T}$ for the ballistic ($\Lambda=0$), $\Lambda=1\mathrm{nm}$ and $\Lambda=2\mathrm{nm}$ cases. The IR scattering increases the width of the LLs to around $4\mathrm{meV}$ for $\Lambda=1\mathrm{nm}$ and $8\mathrm{meV}$ for $\Lambda=2\mathrm{nm}$. The LL profile is found to significantly deviate from the Lorentzian profile which can be explained by the discrete spectrum (a Lorentzian profile is obtained if the retarded self-energy is approximately constant around the given level which is a characteristic of a continuous spectra).

 \subsection{Terminal currents}

 The paramagnetic current flowing through the RTS is obtained by evaluating the trace of the current operator $I_\mathrm{op}(E)$. Following Ref. \onlinecite{datta_mesoscopic}, instead of $I_\mathrm{op}$, we evaluate
\begin{equation}
K_p(E)=\frac{-e}{2\pi \hbar} \left[ \Sigma_p^<(E)G^>(E)-\Sigma_p^>(E)G^<(E)\right], \quad p=\mathrm{E},\mathrm{C},\mathrm{LO}
\end{equation}
which has the same trace as $I_\mathrm{op}(E)$. The symbol $p$ denotes the system with which the current is exchanged\cite{datta_mesoscopic}. For calculating the I-V curves, the emitter (E) and collector (C) currents are needed. The exchange of particles with the phonon bath is described by the $p=\mathrm{LO}$ term.
Utilizing the decoupling of $kn$ coordinates, $K_p(E)$ in the mixed representation reads
\[
K_{p,s_1s_2}(knz_1,z_2;E)=\frac{-e}{2\pi \hbar}\sum_{s_3}\int dz_3 \left[\Sigma^<_{p,s_1s_3}(knz_1,z_3;E)G^>_{s_3s_2}(knz_3,z_2;E)\right.
\]
\begin{equation} \label{eq_kop}
\left.-\Sigma^>_{p,s_1s_3}(knz_1,z_3;E)G^<_{s_3s_2}(knz_3,z_2;E)\right].
\end{equation}
The diagonal elements $K_{p,\sigma\sigma}(\boldsymbol{\rho} z,\boldsymbol{\rho} z;E)$ in the real-space representation are found from a transformation given in Eq. (\ref{eq_GF_realspace}). Finally, the spin and energy resolved current density is obtained by integrating over $z$
\begin{equation} \label{eq_i_from_kop}
i_{p,\sigma}(E)=\int d z K_{p,\sigma\sigma}(\boldsymbol{\rho} z,\boldsymbol{\rho} z;E).
\end{equation}
The total currents are obtained by integrating over all the energies
\begin{equation}
  i_{p,\sigma}=\int dE i_{p,\sigma}(E).
  \end{equation}
Charge and spin conservation implies $i_{\mathrm{E},\sigma}=i_{\mathrm{C},\sigma}$, so if only the total currents are required, either of the two can be calculated. However, the emitter and collector currents have a different energy distribution when dissipation (due to inelastic scattering) is present which, as shown below, provides information on what happens with electrons on their way through the RTS.

We discuss the details of electron transport through RTS on the example of $B=5\mathrm{T}$. While the choice of a particular value of $B$ is not critical, this one has been taken based on several criteria. Firstly, it is large enough ($\hbar \omega_\mathrm{C}=23.9\mathrm{meV}$) that only the lowest LL subband of the contacts is populated, which makes the discussion simpler. This is because the electrochemical potential $\mu$ of both contacts at $B=0$ is assumed to be $\mu=20\mathrm{meV}$, while at $B=5\mathrm{T}$ it is found to be around $\mu=18.2\mathrm{meV}$ (also see the Fermi level variation in Fig. \ref{fig_DOS_phonon}). The dependence of $\mu$ on $B$ corresponds to a three dimensional system (the contacts are also assumed to be exposed to the magnetic field). Secondly, the LL spacing is larger than the level broadening, so the features in the I-V curves can be observed clearly. Thirdly, the lowest magnetophonon resonance (defined by $E_\mathrm{LO}=\hbar \omega_\mathrm{C}$) is close so we can discuss how it is manifested in the RTS currents. Finally, the Zeeman energy $\Delta_\mathrm{B}=-1.45\mathrm{meV}$ at this field determines the spin-splitting (of the lowest LLs) while RSOI is practically negligible. The consequence is that the LLs are polarized approximately either spin-up ($\vert kn +\rangle\approx \vert kn \downarrow\rangle$) or spin-down ($\vert k (n+1)-\rangle \approx \vert k n \uparrow\rangle$) so the transport of the two spin polarizations is practically decoupled. We will consider the spin-up current which is dominated by the $\vert kn-\rangle$ states, the admixing with $\vert kn + \rangle$ states being negligible. For this reason, in explanations we mention only the $\vert kn-\rangle$ states, even though the mixing is taken into account in numerical calculations.

  \begin{figure}
\includegraphics[width=10cm]{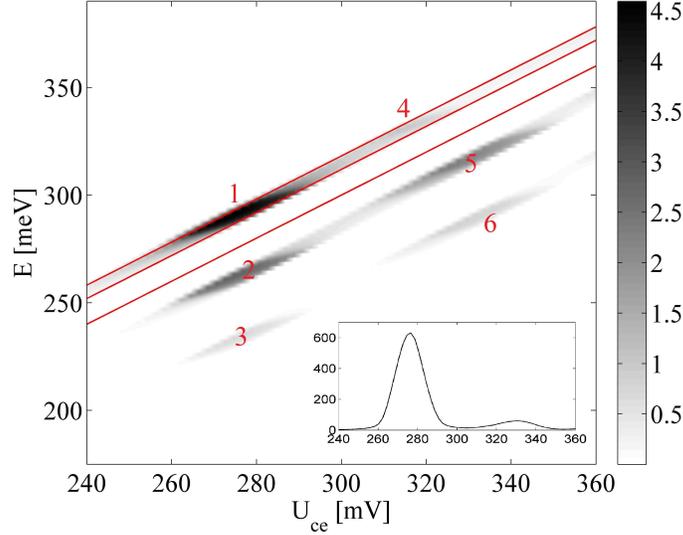}
\caption{\label{fig_iCe_lowV} The energy resolved spin-up collector current $i_{\mathrm{C},\uparrow}(E)$ evolution with $U_\mathrm{ce}$, low voltage range. To emphasize the low value peaks, the map shows $\log\left[1+100\times i_{\mathrm{C},\uparrow}(E)/I_\mathrm{max}\right]$, where $I_\mathrm{max}=114.6\mathrm{A}/\mathrm{cm}^2\mathrm{meV}$ is the highest value of $i_{\mathrm{C},\uparrow}(E)$ in the investigated range. The inset shows the total current $i_{\mathrm{C},\uparrow}$ where the horizontal axis is $U_\mathrm{ce}$ in units of mV and the vertical axis is given in units of $A/\mathrm{cm}^2$. The enumerated peaks are explained in the text.}
\end{figure}

  The transport of electrons across the RTS is studied by considering the energy resolved collector spin-up current $i_{\mathrm{C},\uparrow}(E)$ shown in Fig. \ref{fig_iCe_lowV}. Red lines show the position of the emitter band edge $U_\mathrm{e}$ (lower), LL subband bottom $U_\mathrm{e}+\hbar\omega_\mathrm{C}/2$ (middle) and the top of the emitter Fermi sea, $U_\mathrm{e}+\mu$. The zero values between $U_\mathrm{e}$ and $U_\mathrm{e}+\hbar\omega_\mathrm{C}/2$ are due to that fact that the emitter DOS is zero in this energy range. We do not consider the emitter current $i_{\mathrm{E},\uparrow}(E)$ separately, as it offers little information. This is because we know by default that all the electrons that leave the emitter have energies between $U_\mathrm{e}+\hbar\omega_\mathrm{C}/2$ and $U_\mathrm{e}+\mu$. The total current is shown in the inset. For $U_\mathrm{ce}<240\mathrm{mV}$ the currents are practically zero because the quasi-bound states are above the emitter Fermi sea.

  The peaks in $i_{\mathrm{C},\uparrow}(E)$ labelled by $\mathbf{1}$, $\mathbf{2}$ and $\mathbf{3}$ for $U_\mathrm{ce}$ between 260 and 290 form the main peak of the current. In this range of $U_\mathrm{ce}$ the emitter Fermi sea is aligned with the lowest lying $\vert k1-\rangle$ quasi-bound state, see Fig. \ref{fig_DOS_phonon}. $\mathbf{1}$ denotes the electrons that preserved their energy while going through the device. The majority of this current represents the electrons that have coherently tunneled through the RTS while a small part of them have been elastically scattered into any of the LL subbands in the collector. $\mathbf{2}$ is exactly $E_\mathrm{LO}$ below $\mathbf{1}$, implying that it represents electrons that have entered the well, emitted one LO phonon and got scattered in the collector. Similarly, $\mathbf{3}$ is the current component of resonant transmission with emitting two LO phonons.

  The second peak of the total current seen in the inset of Fig. \ref{fig_iCe_lowV} occurs when $U_\mathrm{ce}$ is increased enough so that the $\vert k 2-\rangle$ quasi-bound level and the phonon replica of $\vert k 1-\rangle$, see Fig. \ref{fig_DOS_phonon} (b), are aligned with the emitter Fermi sea. The peak labelled by $\mathbf{4}$ represents resonant transmission mediated by IR scattering of $\vert k 1-\rangle$ levels into the $\vert k2-\rangle$ levels in the well. It is seen as a shoulder of the peak in the inset. In agreement with Fig. \ref{fig_DOS_phonon} (b), $\mathbf{4}$ occurs for slightly lower $U_\mathrm{ce}$ than $\mathbf{5}$, which represents the resonant transmission via the phonon replica of $\vert k 1-\rangle$. With increasing $B$, $\mathbf{4}$ and $\mathbf{5}$ are found to move in opposite direction, $\mathbf{4}$ moving towards higher $U_\mathrm{ce}$. The anticrossing of the LLs is seen as a disappearance of $\mathbf{4}$ at the magnetophonon resonance (around 6T) and reappearance at higher $U_\mathrm{ce}$, so that $\mathbf{4}$ and $\mathbf{5}$ never occur at the same value of $U_\mathrm{ce}$. Resonant transmission with two LO phonon emission is again observed as $\mathbf{6}$. The resonant enhancement of this process is a consequence of the increased population in the $\vert k1-\rangle$ level into which electrons get by emitting the first LO phonon. Some of the electrons then pass coherently to the collector ($\mathbf{5}$) while the remaining emit an additional phonon.

 \begin{figure}
\includegraphics[width=10cm]{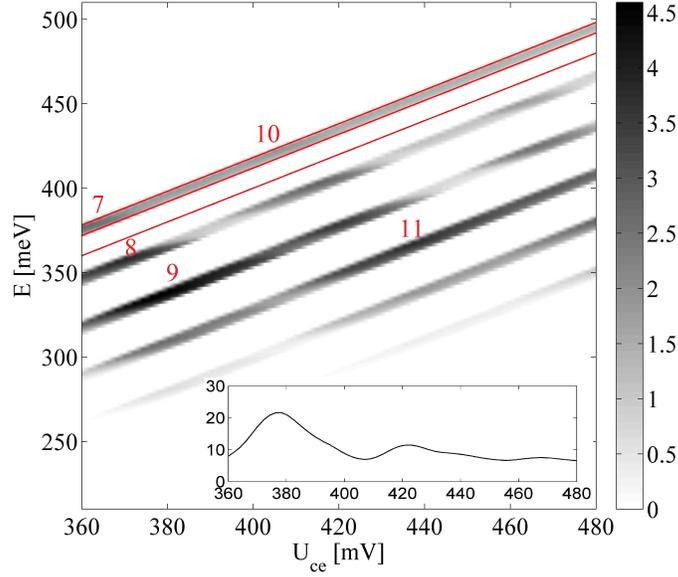}
\caption{\label{fig_icE_highV} The energy resolved spin-up collector current $i_{\mathrm{C},\uparrow}(E)$ evolution with $U_\mathrm{ce}$, the high voltage range. The map shows $\log\left[1+100\times i_{\mathrm{C},\uparrow}(E)/I_\mathrm{max}\right]$, with $I_\mathrm{max}=3.2\mathrm{A}/\mathrm{cm}^2\mathrm{meV}$. The inset is the total current $i_{\mathrm{C},\uparrow}$ where the horizontal axis is $U_\mathrm{ce}$ in units of mV and the vertical axis is given in units of $A/\mathrm{cm}^2$. The enumerated peaks are explained in the text.}
\end{figure}

 The same situation but for $U_\mathrm{ce}$ ranging from 360 to 480mV is shown in Fig. \ref{fig_icE_highV}. The most apparent difference compared to the lower voltage range is the appearance of higher order (up to five) LO phonon emission processes. This is because with increasing $U_\mathrm{ce}$ more LLs in the well are available to scatter into by emitting an LO phonon. The peak labelled by $\mathbf{7}$ occurs when $\vert k3-\rangle$ is alligned with the emitter Fermi sea. At slightly higher $U_\mathrm{ce}$ appear $\mathbf{8}$ and $\mathbf{9}$ corresponding to the first replica of $\vert k2-\rangle$ and second replica of $\vert k1-\rangle$, respectively. In the high voltage range an important difference between the elastic and inelastic currents is seen: the elastic current appears almost as a structureless stripe in the $U_\mathrm{e}+\hbar\omega_\mathrm{C}/2$ to $U_\mathrm{e}+\mu$ energy range. The peak labelled as $\mathbf{10}$ corresponding to IR scattering via the $\vert k 4-\rangle$ state is hardly visible. This is probably due to the gradual shifting of the LLs in the well towards the collector, which decreases their interaction with emitter states via the IR of the emitter barrier. On the contrast, the inelastic current comprises many peaks, each representing the alignment of a phonon replica with the emitter Fermi sea. $\mathbf{11}$ corresponds to the third phonon replica of the $\vert k1-\rangle$ state. When considering the entire range of voltages, it is seen that the peaks $\mathbf{5}$, $\mathbf{9}$ and $\mathbf{11}$, corresponding to the first, second and third replica of $\vert k1-\rangle$ are the most pronounced in the inelastic current. It seems counterintuitive that $\mathbf{9}$ and $\mathbf{11}$ should be higher than some of the peaks with less phonons emitted, because the two-phonon or three-phonon emission processes are weaker that the one-phonon emission process. This is also seen in Fig. \ref{fig_DOS_phonon} (b) where the second replica of $\vert k1-\rangle$ clearly has a lower DOS than the first replica. While the results indicate that second and third order processes make a difference in calculations, it should be noted that the reason that the two- and three-phonon peaks in $\mathbf{9}$ and $\mathbf{11}$ are more intense then the lower order peaks is the sequential emission of one phonon at a time. This is possible, even though the magnetophonon resonance $E_\mathrm{LO}=\hbar \omega_\mathrm{C}$ is not exactly achieved, due to the significant level broadening caused by IR scattering.

   \begin{figure}
\includegraphics[width=10cm]{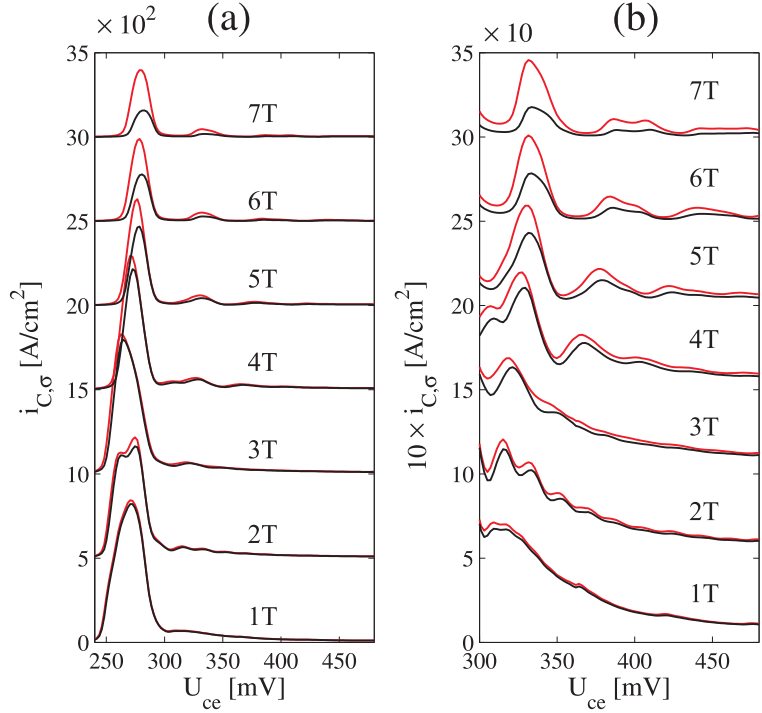}
\caption{\label{fig_currents_vs_B} (a) I-V curves for with $B$ changed as a parameter. In (b) the I-V curves for $U_\mathrm{ce}>300\mathrm{mV}$ are shown magnified by a factor of 10. The curves in (a) and (b) are vertically displaced by $500\mathrm{A}/\mathrm{cm}^2$ and $50\mathrm{A}/\mathrm{cm}^2$, respectively. The red lines represent the spin-up, while the black lines are the spin-down currents.}
\end{figure}

 The effect of varying the magnetic field on the total currents is shown in Fig. \ref{fig_currents_vs_B}. There is very little difference between the $B=1\mathrm{T}$ and $B=0\mathrm{T}$ curves, the latter not being shown in Fig. \ref{fig_currents_vs_B} (see, for example, Fig. 4 in Ref. \onlinecite{isic_jap_2010}). The is because the broadening of quasibound states is larger than $\hbar\omega_\mathrm{C}=4.8\mathrm{meV}$. For $B=2\mathrm{T}$, the emitter chemical potential is around $20\mathrm{meV}$ so two LL subbands are occupied, leading to the structure in the main peak (the onset of the coherent $\vert k 2-\rangle$ current is around $U_\mathrm{ce}=265\mathrm{mV}$), and the satellite peaks - the two peaks in the $310-340\mathrm{mV}$ range are the first replicas of $\vert k 1-\rangle$ and $\vert k 2-\rangle$. At $B=3\mathrm{T}$, $\mu=21.6\mathrm{meV}$, while $3/2\times \hbar\omega_\mathrm{C}=21.5\mathrm{meV}$, so the second emitter LL subband is almost cut off. The coherent current from the $\vert k 2-\rangle$ emitter subband participates in the current but cannot be observed in the I-V curve. From $B=4\mathrm{T}$ and above, only the lowest LL emitter subband is occupied. The peaks seen in the I-V curves occur when the alignment between either a higher LL or a phonon replica with the emitter Fermi sea is achieved. The evolution of the peaks towards higher values of $U_\mathrm{ec}$ reflects the evolution of the DOS with $B$, as discussed on the example of the $B=5\mathrm{T}$ case.

    \begin{figure}
\includegraphics[width=10cm]{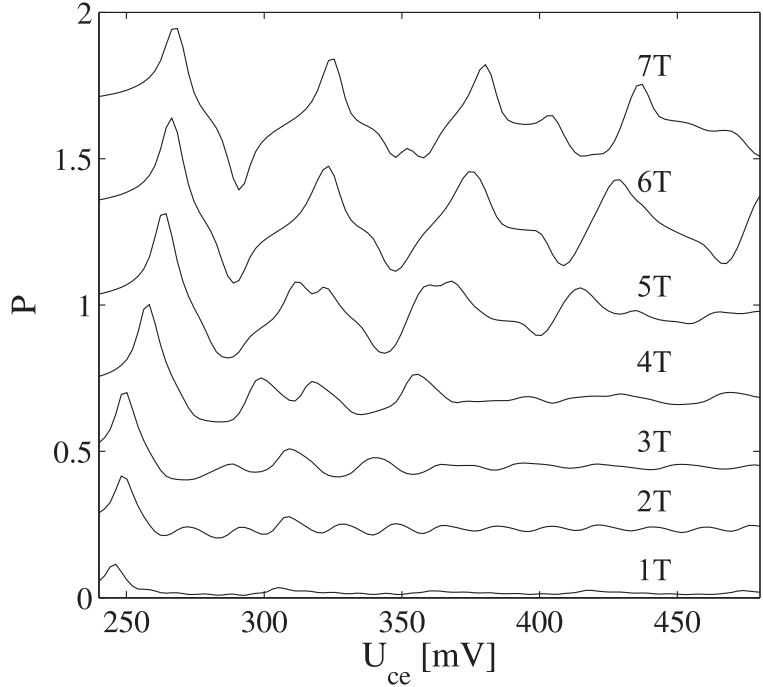}
\caption{\label{fig_pol_vs_B} Spin polarization $P$ for various $B$. The curves are vertically displaced by $0.2$.}
\end{figure}

 The spin polarization $P$ of the current, defined as
 \begin{equation}
 P=\frac{i_{\mathrm{C},\uparrow}-i_{\mathrm{C},\downarrow}}{i_{\mathrm{C},\uparrow}+i_{\mathrm{C},\downarrow}},
 \end{equation}
 with the magnetic field varied as a parameter is plotted in Fig. \ref{fig_pol_vs_B}. The highest value of $P$ is always reached at the main peak but significant values occur at satellite peaks as well. To explain this, we argue that the value of $P$ is determined by two factors: (1) the magnetization of the emitter Fermi sea which becomes significant once only the lowest LL subband is populated and (2) the spin splitting of quasi-bound LLs in the well. Obviously, the magnetization is also determined by the spin splitting (but that of the emitter states). However, the distinction is important because (1) and (2) represent two mechanisms leading to nonzero $P$: the former influences the electron supply while the latter determines how well will the RTS itself filter the electrons based on their spin. $P$ is higher at the main peak because most of the current is coherent and the quasi-bound LLs involved are narrower. However, the fact that the structure in $P$ reflects so clearly the scattering processes means that (2) is significant even at the satellite peaks.

\section{Summary}

A comprehensive model for quantum transport in RTSs in a perpendicular magnetic field has been described. Effects like the formation of magnetopolarons and IR induced LL broadening  are accurately reflected in the evolution of the I-V curves with changing the magnetic field. Using effective parameters that realistically reflect the InAs bandstructure, a beating pattern in the DOS has been obtained even with large IR scattering induced broadening, which is in agreement with experimental findings that the RSOI constant can be estimated from SdH measurements. It has been found that the I-V curves are not sensitive to RSOI, and that the spin polarization is dominated by the Zeeman effect. The spin polarization of the current is found to be significant for magnetic fields of few Tesla even at satellite peaks, suggesting that its measurement might yield additional information in magnetospectroscopic studies.

\begin{acknowledgments}
 Authors acknowledge the support of NATO Collaborative Linkage Grant (Reference No. CBP.EAP.CLG 983316). G.I. is grateful for the support from ORSAS (UK), University of Leeds, the School of Electronic and Electrical Engineering, and the Serbian Ministry of Science Project No. OI171005. V.M. and J.R. are grateful for support from the Serbian Ministry of Science Project No. III45010.
\end{acknowledgments}

\appendix

\section{Correlation function for IR scattering} \label{sec_IR_corr_funcs}

 The quantity used to describe interface roughness is the interface fluctuation $\Delta_\mathrm{IR}(\mathbf{r})$, i.e. the length by which the interface is displaced from the average value. By definition, the average (over $\mathbf{r}$) value of $\Delta_\mathrm{IR}(\mathbf{r})$ is zero. Typical values of $\Delta_\mathrm{IR}$ are of the order of one monolayer, i.e. few angstroms (see e.g. Ref. \onlinecite{sakaki_interface_roughness_scatt_in_algaas_qw_1987}), therefore the $z$-dependence of the scattering potential caused by the roughness, $V_\mathrm{IR}(\mathbf{r})$, can be taken as a delta function and $\Delta_\mathrm{IR}$ considered to be a function of only the lateral coordinates $\Delta_\mathrm{IR}(\rho)$
 \begin{equation} \label{eq_interf_scatt_pot}
 V_\mathrm{IR}(\mathbf{r})=\delta(z-z_\mathrm{I})V_\mathrm{b}\Delta_\mathrm{IR}(\boldsymbol{\rho}),
 \end{equation}
 where $V_\mathrm{b}$ represents the conduction band discontinuity at the heterointerface.

 If the interaction between electrons and a scatterer is described by the Hamiltonian
 \begin{equation} \label{eq_app_int_h}
 H'=\int d\mathbf{r}\Psi^\dag(\mathbf{r};t) V(\mathbf{r};t)\Psi(\mathbf{r};t),
 \end{equation}
 where $V(\mathbf{r};t)$ can be a static external potential (in which case $t$ is redundant) or involve dynamic degrees of freedom as in case of phonons, the scatterer correlation functions are defined as \cite{datta_kinetic_eq_1990}
 \begin{equation}
 D^>(\mathbf{r}_1,\mathbf{r}_2;t_1,t_2)=\left\langle V(\mathbf{r}_1;t_1)V(\mathbf{r}_2;t_2)\right\rangle, \quad
 D^<(\mathbf{r}_1,\mathbf{r}_2;t_1,t_2)=\left\langle V(\mathbf{r}_2;t_2)V(\mathbf{r}_1;t_1)\right\rangle.
 \end{equation}
 The brackets $\langle ... \rangle$ imply the statistical average which is the configurational average in case of the IR scattering or the ensemble average in case of phonons.

 For IR scattering the interaction potential $V_\mathrm{IR}(\mathbf{r})$ is static so the lesser and greater correlation functions are equal. After transforming to the energy domain, we find
 \begin{equation}
D_\mathrm{IR}(\mathbf{r}_1,\mathbf{r}_2;E)=2\pi\hbar\delta(E)\delta(z_1-z_I) \delta(z_1-z_2) V_\mathrm{b}^2 \langle \Delta_\mathrm{IR}(\boldsymbol{\rho}_1) \Delta_\mathrm{IR}(\boldsymbol{\rho}_2)\rangle.
\end{equation}
The quantity $\langle \Delta_\mathrm{IR}(\boldsymbol{\rho}_1) \Delta_\mathrm{IR}(\boldsymbol{\rho}_2)\rangle$ is called the IR autocorrelation function in the literature and usually assumed to be of the Gaussian form \cite{ando_2deg_1982,sakaki_interface_roughness_scatt_in_algaas_qw_1987}
\begin{equation}
\langle \Delta_\mathrm{IR}(\boldsymbol{\rho}_1) \Delta_\mathrm{IR}(\boldsymbol{\rho}_2)\rangle_\mathrm{av}=\Delta e^{-(\boldsymbol{\rho}_1-\boldsymbol{\rho}_2)^2/\Lambda^2},
\end{equation}
where $\Delta$ is referred to as the roughness height and $\Lambda$ as the roughness lateral length.  The values reported in Ref. \onlinecite{sakaki_interface_roughness_scatt_in_algaas_qw_1987} are $\Delta=3-5\AA$ and $\Lambda=50-70\AA$. In problems where $\Lambda$ is small compared to other relevant lengths, the Gaussian may be approximated by a delta function, so
\begin{equation}
D(\mathbf{r}_1\mathbf{r}_2;E)=2\pi\hbar\delta(E)\delta(z_1-z_I)\delta(z_1-z_2)\delta(\boldsymbol{\rho}_1-\boldsymbol{\rho}_2) V_\mathrm{b}^2 \Delta^2 \pi \Lambda^2.
\end{equation}
When a magnetic field is perpendicular to the interface the magnetic length $L=\sqrt{\hbar/eB}$ is the relevant quantity. As it becomes smaller than $10\mathrm{nm}$ for $B>6.2\mathrm{T}$, we find that the delta approximation cannot be considered as strictly accurate in the investigated range of magnetic fields (up to $10\mathrm{T}$) but that it can still be expected to give qualitatively correct results.

\section{LO phonon correlation function} \label{sec_LO_corr_funcs}

For polar coupling to bulk LO phonons (Fr\"{o}lich interaction), the term $V(\mathbf{r};t)$ in Eq. (\ref{eq_app_int_h}) is given by\cite{mahan_1990}
\begin{equation}
V(\mathbf{r};t)=\sum_\mathbf{q} M_\mathbf{q}e^{-i\mathbf{qr}}A_\mathbf{q}(t), \quad
M_\mathbf{q}=\frac{1}{q}
\sqrt{\frac{e^2 E_\mathrm{LO}}{2\Omega \varepsilon_0}
\left(\frac{1}{\varepsilon(\infty)}-
\frac{1}{\varepsilon(0)}\right)},
 \end{equation}
 where $A_\mathbf{q}(t)=a_\mathbf{q}(t)+a^\dag_\mathbf{q}(t)$ is the phonon operator while $a_\mathbf{q}(t)$ and $a^\dag_\mathbf{q}(t)$ are the phonon destruction and creation operators, respectively. Assuming that the phonons are in a thermodynamic equilibrium and neglecting the LO phonon dispersion, $E_\mathbf{q}\equiv E_\mathrm{LO}$, we find
 \begin{equation}
\langle A_{\mathbf{q}_1}(t_1) A_{\mathbf{q}_2}(t_2)\rangle=
\delta_{\mathbf{q}_1,-\mathbf{q}_2}\left\{e^{-iE_\mathrm{LO}(t_1-t_2)/\hbar}\left[n_\mathrm{B}(E_\mathrm{LO})+1\right]+
e^{iE_\mathrm{LO}(t_1-t_2)/\hbar} n_\mathrm{B}(E_\mathrm{LO})\right\}.
\end{equation}
At low temperatures such that $k_\mathrm{B}T\ll E_\mathrm{LO}$, the Bose-Einstein occupation factor is very small, $n_\mathrm{B}(E_\mathrm{LO})\ll1$, so it can be set to zero (only spontaneous phonon emission is considered). Upon switching to the energy domain and summing up over $\mathbf{q}$, the LO phonon correlation functions are found as
 \begin{equation} \label{eq_app_DLO_exact}
 D_\mathrm{LO}^\gtrless(\mathbf{r}_1,\mathbf{r}_2;\omega)=\delta(\omega\mp \omega_\mathrm{LO})\frac{C_\mathrm{LO}}{\vert \mathbf{r}_1-\mathbf{r}_2\vert}, \quad
 C_\mathrm{LO}=\frac{e^2\hbar\omega_\mathrm{LO}}{4\varepsilon_0}\left(\frac{1}{\varepsilon(\infty)}-\frac{1}{\varepsilon(0)}\right).
 \end{equation}

 In the main text, the correlation functions are approximated by
  \begin{equation} \label{eq_DLO_local}
 D_\mathrm{\delta LO}^\gtrless(\mathbf{r}_1,\mathbf{r}_2;\omega)=2\pi \delta(\omega\mp\omega_\mathrm{LO}) U_\mathrm{LO}^2 \delta(\mathbf{r}_1-\mathbf{r}_2).
\end{equation}
 In contrast to the case of IR scattering where the scatterer correlation is a short range Gaussian function, approximating the long range $1/r$ correlation by a delta function may appear bizarre. The first issue to be resolved when the $1/r$ correlation is used is finding the relevant length scale. In extended systems, such as the bulk, the scale is set by the screening length. As the problem of electron-phonon interaction in a RTS is analogous to the problem of electron-phonon interaction in a quantum well, we may consider the latter, leaving screening aside, and conclude that for our purposes the scale is set by the quantum well width $L_z$ and the magnetic length $L$. But since $L_z$ is the extension of the entire investigated system, approximating $D^\gtrless_\mathrm{LO}$ by $D^\gtrless_{\delta\mathrm{LO}}$ is obviously flawed in a fundamental way. However, we will now show that the scattering strength $U_\mathrm{LO}$ can be chosen so that $D^\gtrless_{\delta\mathrm{LO}}$ gives scattering rates for electronic states in a quantum well that are close to scattering rates obtained using $D^\gtrless_\mathrm{LO}$.

 We consider an infinitely deep quantum well of width $L_z$ in a perpendicular magnetic field. $k$, $n$ and the space coordinates are the same as in the main text. As RSOI and the Zeeman effect are irrelevant, we neglect them. The eigenkets are $\vert kna\rangle$ where $a$ denotes the quantum number for motion along the $z$ axis. As the aim is only to compare the two models, it is sufficient to consider FBA without self-consistency. The Green's functions below are, therefore, the non-interacting Green's functions with only diagonal elements in $kna$, labeled by $G^\gtrless(na;E)$.  The self-energies for the delta model are evaluated straightforwardly as
 \begin{equation}\label{eq_app_sigmadlo}
 \Sigma^\gtrless_{\delta\mathrm{LO}}(k_1n_1a_1,k_2n_2a_2;E)=\delta_{k_1k_2}\delta_{n_1n_2}\frac{U_\mathrm{LO}^2}{2\pi L^2} \sum_{n_3a_3}G^\gtrless(n_3a_3;E\mp E_\mathrm{LO})\int_0^{L_z} dz
 \left\vert \langle z\vert a_2\rangle\right\vert^2\left\vert \langle z\vert a_3\rangle\right\vert^2.
 \end{equation}

 \begin{figure}
\includegraphics[width=10cm]{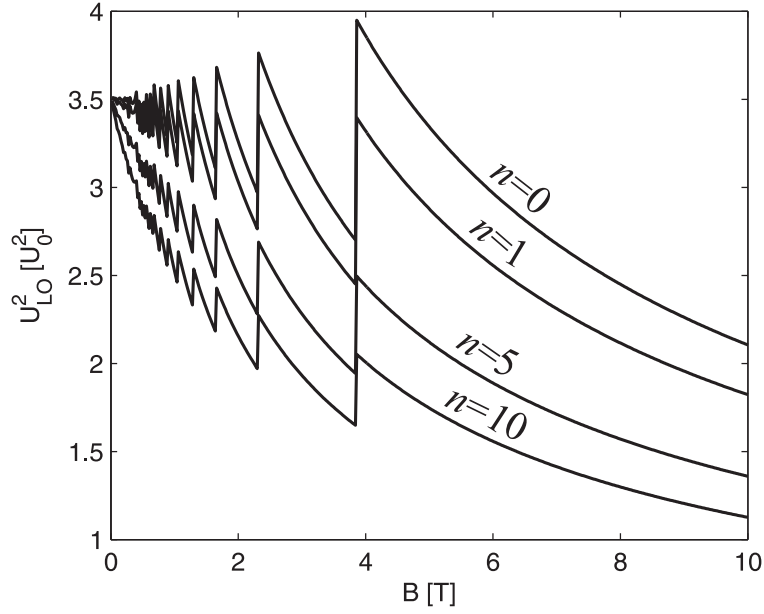}
\caption{\label{fig_alphavar_vs_B} LO phonon scattering strength $U_\mathrm{LO}^2$ as a function of the magnetic field, calculated according to Eq. (\ref{eq_app_ULO}) and expressed in units of $U_0^2$.}
\end{figure}

The self-energies for $D^\gtrless_\mathrm{LO}$ are not diagonal in $kn$, but for the first order effect we need to consider only the diagonal elements labelled by $\Sigma^\gtrless_\mathrm{LO}(kna;E)$
 \begin{equation}\label{eq_app_sigmalo}
 \Sigma^\gtrless_\mathrm{LO}(kna;E)=\sum_{n_1a_1}G^\gtrless(n_1a_1;E\mp E_\mathrm{LO})\frac{\Omega}{(2\pi)^3}\int d\mathbf{q}
  M_\mathbf{q}^2 \vert F(q_z,a_1,a)\vert^2 \vert H(q_\parallel,n_1,n)\vert^2,
 \end{equation}
 where
 \begin{equation}
F(q_z,a_1,a)=\int dz \langle a\vert z\rangle e^{-iq_zz}\langle z \vert a_1\rangle,
\end{equation}
and
\begin{equation} \label{eq_Hsq_final}
 \vert H(q_\parallel,n_1,n) \vert^2=e^{-\xi^2/2}\frac{n!}{n_1!}\left(\frac{\xi^2}{2}\right)^{n_1-n}\left| L_{n}^{n_1-n}\left(\frac{\xi^2}{2}\right)\right|^2,
 \quad n_1\geq n, \quad \xi=q_\parallel L.
 \end{equation}
 In these equations $\mathbf{q}$ is the phonon wavevector with $\mathbf{q}_\parallel$ and $q_z$ being its in-plane and perpendicular components while $L^m_n$ is the generalized Laguerre polynomial. For more details on evaluating the electron-LO phonon  matrix elements\cite{barker_magnetophonon_effect_1972} and scattering rates in a 2DEG see Refs. \onlinecite{becker_lo_phonon_ll_2004,savic_transport_in_magn_qcl_2006,savic_density_matrix_theory_qcl_magnetic_field_2007}. Assuming that only the lowest lying $a=1$ subband in an infinitely deep quantum well is populated (as is the case with the RTS investigated in the main text) with $\langle z\vert 1\rangle=\sqrt{2/L_z}\sin(\pi z/L_z)$, the self-energies given by Eqs. (\ref{eq_app_sigmadlo}) and (\ref{eq_app_sigmalo}) are seen to be equal if
 \begin{equation} \label{eq_app_ULO}
 U_\mathrm{LO}^2=\frac{L^2L_z^2e^2E_\mathrm{LO}}{12\pi^2 \varepsilon_0}\left(\frac{1}{\varepsilon(\infty)}-\frac{1}{\varepsilon(0)}\right)Q(n_1,n_2,B),
 \end{equation}
 \begin{equation}
 Q(n_1,n_2,B)=\int d\mathbf{q}\frac{\vert F(q_z,1,1)\vert^2 \vert H(q_\parallel,n_1,n_2)\vert^2}{q^2}.
 \end{equation}
 The value of $U_\mathrm{LO}^2$ given by Eq. (\ref{eq_app_ULO}) ensures that the scattering rate from LL $n_1$ to the lower lying $n_2$ due to LO phonon emission obtained assuming the delta correlation $D_\mathrm{\delta LO}^\gtrless$ is the same as the one obtained from the exact $D_\mathrm{LO}^\gtrless$. The dependence of $U_\mathrm{LO}^2$ on $B$ and  LLs involved in scattering is analyzed by noting that only $n_1,n_2$ pairs which differ in energy by $E_\mathrm{LO}$ are relevant. Therefore, we fix the lower LL index $n_2$ in Eq. (\ref{eq_app_ULO}) and numerically integrate $Q(n_1,n_2,B)$ to obtain the scattering strength as a function of $B$, while choosing the upper LL index $n_1$ so that $n_1=n_2+[E_\mathrm{LO}/\hbar\omega_\mathrm{C}]$, where the square brackets $[u]$ denote the integer closest to $u$. The calculated $U_\mathrm{LO}^2$ for $L_z=6\mathrm{nm}$ are shown in Fig. \ref{fig_alphavar_vs_B} in units of $U_0^2=1\mathrm{meV}\,\mathrm{nm}/D_{\mathrm{2DEG}}$. These results show that the scattering strengths that should be used to accurately describe the strength of various inter-LL transitions in the investigated range of magnetic fields are well within the order of magnitude with each other. Consequently, a $D_\mathrm{\delta LO}^\gtrless$ correlation with a fixed (i.e. independent of $n_1,n_2$ and $B$) value of $U_\mathrm{LO}^2$ may be expected to give predictions similar to those of a $D_\mathrm{LO}^\gtrless$ model.

\end{document}